\documentclass[]{article}   
\usepackage{graphicx}
\usepackage[utf8]{inputenc}
\usepackage{subcaption}
\usepackage{amsmath}
\usepackage{amssymb}
\usepackage{float}
\usepackage{url}
\usepackage{xcolor}
\usepackage{makecell}
\usepackage{cite}
\usepackage{tabularx}
\usepackage{paralist}
\usepackage{array}
\usepackage{booktabs}
\usepackage{gensymb} 
\usepackage{multirow}
\usepackage{tikz}
\usepackage{adjustbox}
\usepackage{csquotes}

\usepackage[inline,shortlabels]{enumitem}
\usepackage{caption}
\usepackage{tabularx,ragged2e}
\newcolumntype{L}{>{\RaggedRight}X}
\usepackage{makecell}
\usepackage[margin=1in]{geometry}

\usepackage{pgfplots}
\DeclareUnicodeCharacter{2212}{−}
\usepgfplotslibrary{groupplots,dateplot}
\usetikzlibrary{patterns,shapes.arrows}
\pgfplotsset{compat=newest}
\usepackage{pgf}

\newcommand{\luceme}[1]{\textit{#1$_L$}}

\newcommand{\eg}{\textit{e.g.}, }
\newcommand{\etal}{\textit{et al.}}
    
\title{\LARGE\bf HREyes: Design, Development, and Evaluation of a Novel Method \\for AUVs to Communicate Information and Gaze Direction*}

\author{Michael Fulton$^{1}$, Aditya Prabhu$^{2}$, and Junaed Sattar$^{3}$
\thanks{This work was supported by the US National Science Foundation awards IIS-\#184536 \& \#00074041, the MnRI Seed Grant, and the University of Minnesota Undergraduate Research Opportunities Program (UROP).}%
\thanks{The authors are with the Department of Computer Science and Engineering and the Minnesota Robotics Institute,
        University of Minnesota Twin Cities, Minneapolis, MN, USA.
        {\tt\small \{$^{1}$fulto081,$^{2}$prabh079,$^{3}$junaed\}@umn.edu}}%
}

\begin{document}

\maketitle
\thispagestyle{empty}
\pagestyle{empty}

\begin{abstract}
We present the design, development, and evaluation of HREyes: biomimetic communication devices which use light to communicate information and, for the first time, gaze direction from AUVs to humans.
First, we introduce two types of information displays using the HREye devices: active lucemes and ocular lucemes.
\textit{Active lucemes} communicate information explicitly through animations, while \textit{ocular lucemes} communicate gaze direction implicitly by mimicking human eyes.
We present a human study in which our system is compared to the use of an embedded digital display that explicitly communicates information to a diver by displaying text.
Our results demonstrate accurate recognition of active lucemes for trained interactants, limited intuitive understanding of these lucemes for untrained interactants, and relatively accurate perception of gaze direction for all interactants.
The results on active luceme recognition demonstrate more accurate recognition than previous light-based communication systems for AUVs (albeit with different phrase sets).
Additionally, the ocular lucemes we introduce in this work represent the first method for communicating gaze direction from an AUV, a critical aspect of nonverbal communication used in collaborative work.
With readily available hardware as well as open-source and easily re-configurable programming, HREyes can be easily integrated into any AUV with the physical space for the devices and used to communicate effectively with divers in any underwater environment with appropriate visibility.
\end{abstract}
\section{Introduction}
\label{sec:intro}
Advancements in autonomous underwater vehicle (AUV) design have compounded over recent decades to produce new human-scale collaborative AUVs (co-AUVs)~\cite{fujii_development_1996, dudek_aqua_2007, hackbarth_hippocampus_2015, miskovic_caddy-cognitive_2016, edge_design_2020} ideal for work with human partners. 
While larger AUVs have been effectively applied to oceanographic surveys~\cite{wynn_autonomousunderwatervehicles_2014}, wreck exploration~\cite{foley_2005chiosancient_2009}, and underwater construction/infrastructure~\cite{reisenbichler_automatingmbarismidwater_2016} inspection for decades, smaller co-AUVs are being developed for new tasks.
When considering the application of co-AUVs to new tasks such as marine biological survey, trash cleanup, pollution remediation, and complex underwater construction tasks, it is often proposed that human-scale AUVs could work alongside humans, acting as collaborative partners at depth. 
Though co-AUVs are still very much in their infancy, there is a great deal of promise in this approach, which would allow divers and AUVs to benefit from the other's respective strengths.
However, for this collaborative work to be achieved, co-AUVs must be capable of accurate, expressive, and robust communication with their human partners.

While both AUV-to-human (A2H) communication and human-to-AUV (H2A) communication are important capabilities for enabling collaborative underwater work, we concern ourselves with the former: the ability of an AUV to communicate with the divers around it.
Existing methods for A2H communication are insufficiently robust for field deployment, difficult to learn, and limited in the types of information which can be communicated.
The most commonly used method, displaying messages on an integrated digital display~\cite{dudek_aqua_2007,verzijlenberg_swimming_2010,ukai_swimoid_2013,miskovic_caddy-cognitive_2016}, allows for simple communication of high-density information but is intractable when the diver and AUV are any distance apart~\cite{fulton_rcvm-thri_2022}. 
The use of sequences of flashing lights is another commonly proposed methodology~\cite{fujii_development_1996,verzijlenberg_swimming_2010, demarco_underwater_2014, fulton_rcvm-thri_2022}, but the methods evaluated thus far are not intuitive, imposing a significant burden on divers to learn and remember a confusing set of light codes.
Finally, of all the A2H communication methods which have been proposed, (digital displays, light codes, and robot motion gestures~\cite{fulton_rcvm-thri_2022}) none allow communication of robot gaze.
Gaze is a well-established vector of implicit communication, and has been shown to improve coordination and task success in human-robot teams in terrestrial environments~\cite{hoffman_collaboration_2004, breazeal_effects_2005}. 
These problems with existing methods of A2H communication contribute to the current state of underwater human-robot interaction (UHRI): AUV communication is insufficiently robust to be used in anything but the best conditions, not particularly easy to understand, and entirely explicit, with no implicit information channels.

\begin{figure}[t]
    \centering
    \includegraphics[width=\linewidth, trim={5cm 0 0 5cm}, clip]{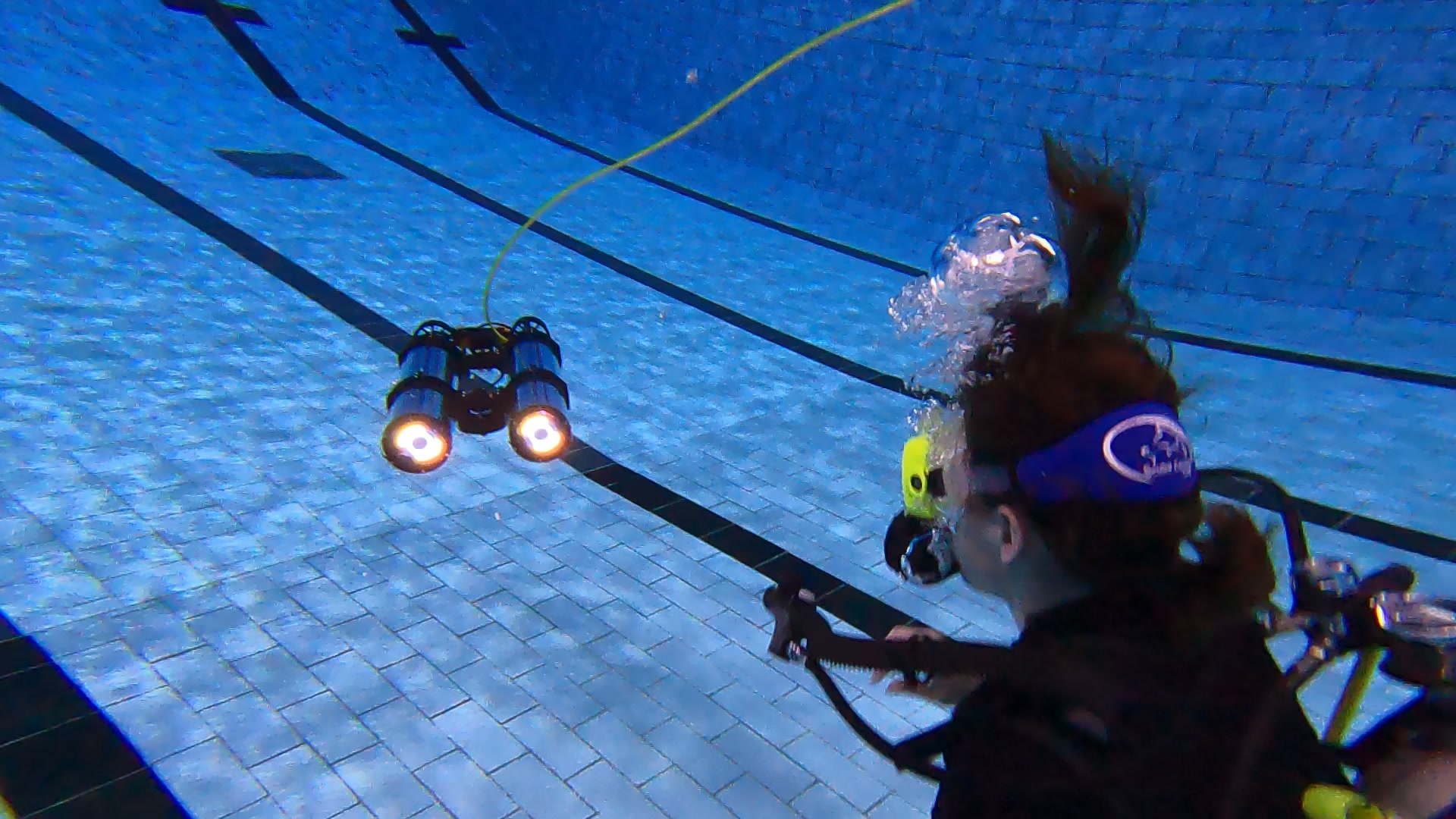}
    \caption{The LoCO AUV informs a diver that it is ready to follow them, using the active luceme \luceme{FollowYou}.}
    \label{fig:intro_figure}
    \vspace{-5mm}
\end{figure}
\begin{figure*}
    \vspace{2mm}
    \centering
    \begin{subfigure}{.24\textwidth}
        \includegraphics[width=\linewidth,trim={0cm 5.25cm 0cm 2.75cm}, clip]{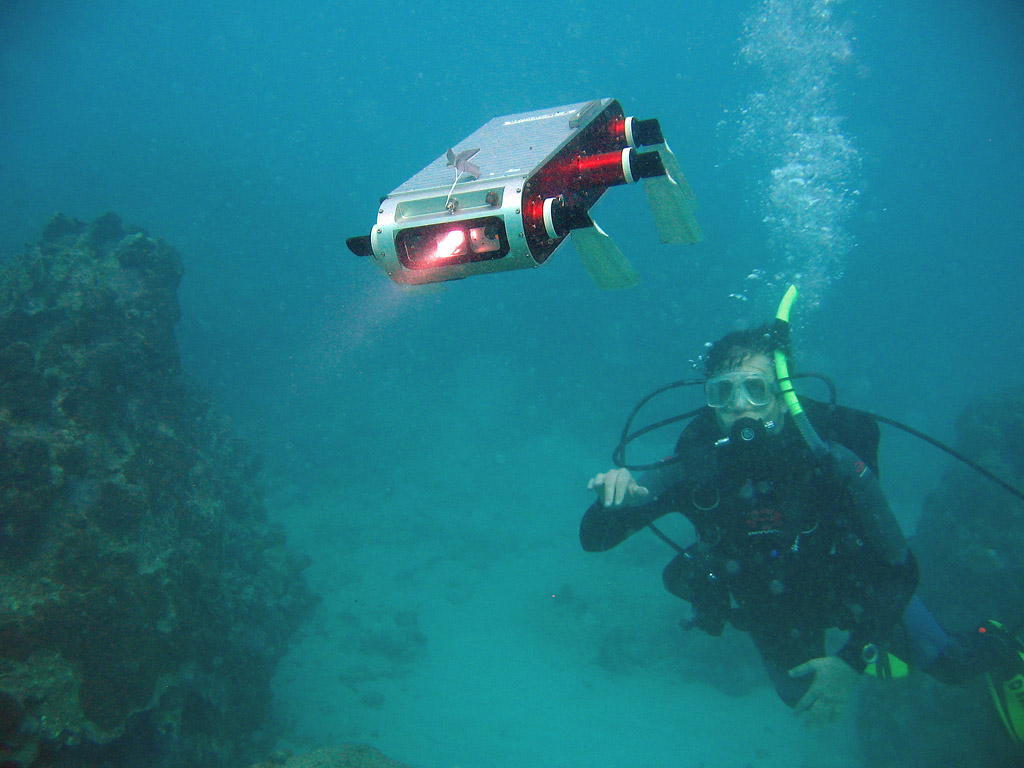}
                \vspace{-5mm}
        \caption{Aqua AUV with 1 light.}
        \label{fig:led_evolution:single}
    \end{subfigure}
    \begin{subfigure}{.24\textwidth}
        \includegraphics[width=\linewidth,trim={1cm 0cm 7cm 0cm}, clip]{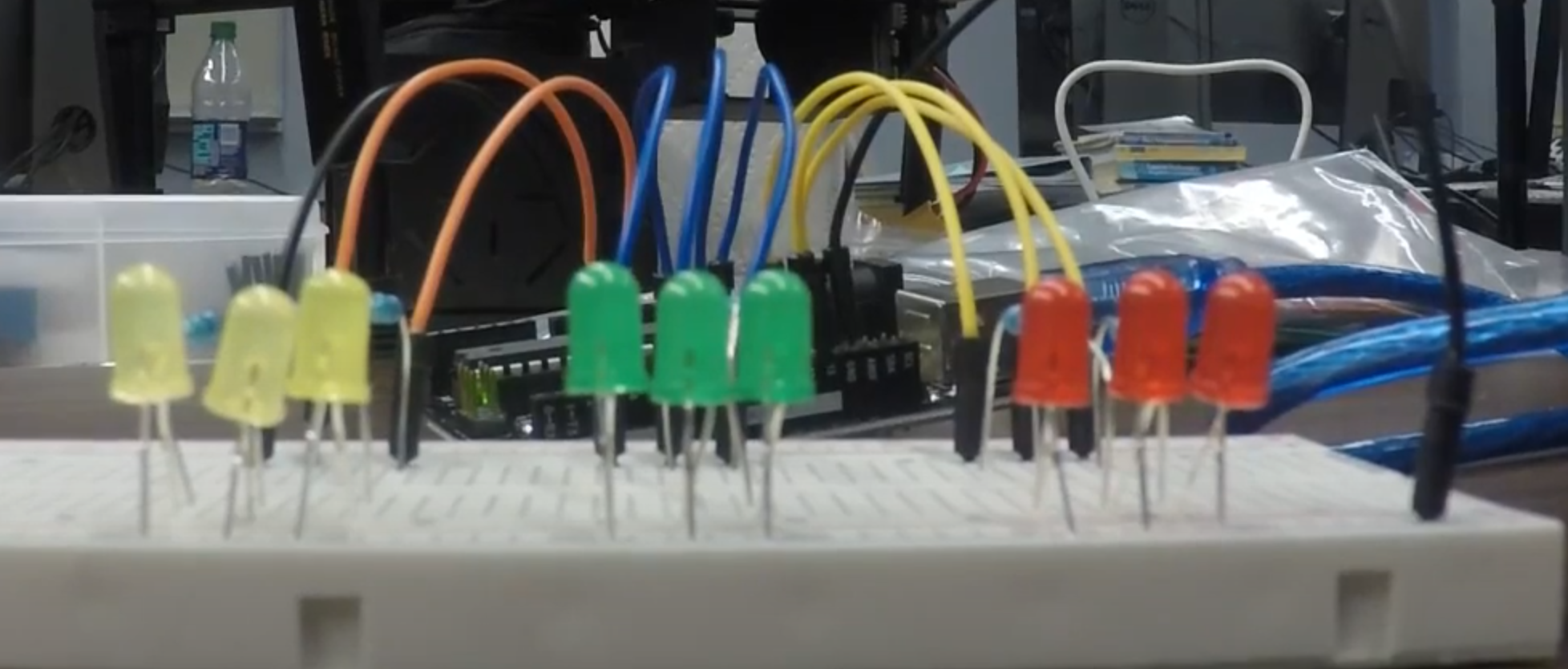}
                \vspace{-5mm}
        \caption{9-LED array~\cite{fulton_rcvm-icra_2019, fulton_rcvm-thri_2022}.}
        \label{fig:led_evolution:nine}
    \end{subfigure}
    \begin{subfigure}{.24\textwidth}
        \includegraphics[width=\linewidth, trim={30cm 14cm 29cm 19.5cm}, clip]{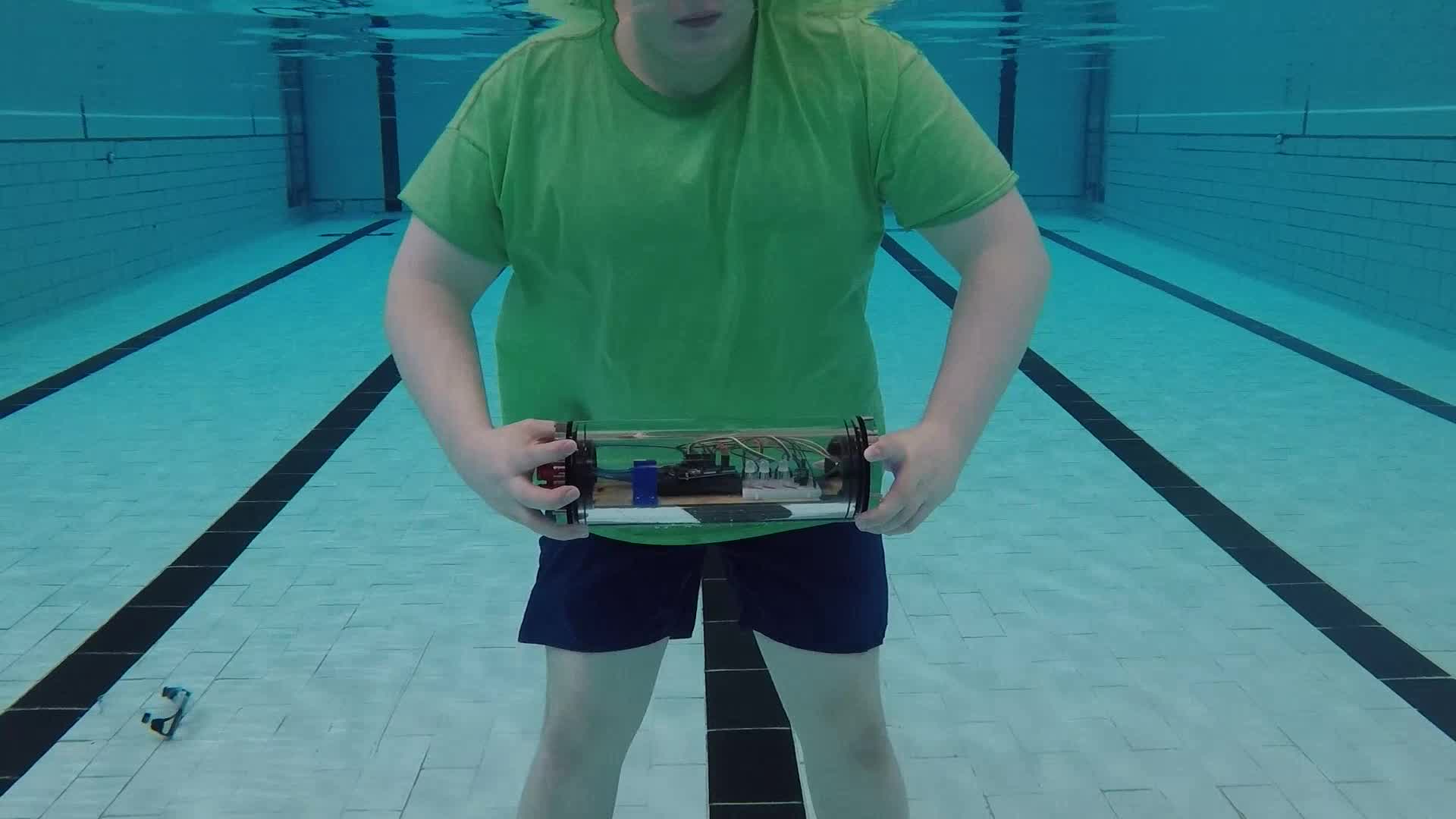}
                \vspace{-5mm}
        \caption{3-LED array~\cite{fulton_rcvm-rss_2022, fulton_rcvm-thri_2022}.}
        \label{fig:led_evolution:three}
    \end{subfigure}
    \begin{subfigure}{.24\textwidth}
        \includegraphics[width=\linewidth, trim={0cm 1.1cm 14cm 3.cm}, clip]{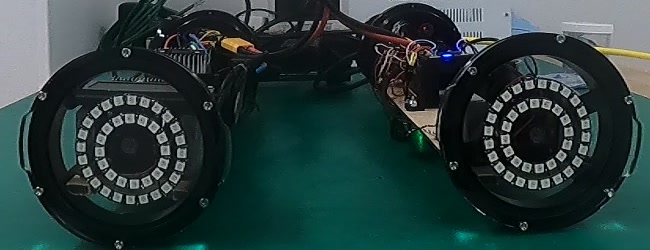}
        \vspace{-5mm}
        \caption{An HREye LED array.}
        \label{fig:led_evolution:hreye}
    \end{subfigure}
    \caption{The evolution of AUV LED communication systems over time. \\\textit{(Figure \ref{fig:led_evolution:single} credit: McGill MRL \& Ioannis Rekleitis)}}
    \label{fig:led_evolution}
    \vspace{-5mm}
\end{figure*}

In this work, we present a communication method that helps to improve A2H communication on all three fronts: the expressive light devices we call \textbf{HREye(s)}. 
HREyes allow for communication at a further distance than digital displays, are more intuitive and expressive than previously proposed light-based communication, and provide a new capability for AUVs: gaze indication.
In the following sections, we first discuss the background of this work with further discussion of existing UHRI methods, brief summaries of research on using light for HRI, and different methods of communicating gaze in terrestrial robots.
Following this, we present the design of the HREye devices, their evolution from earlier light-based methods we have proposed, and the software used to control them.
The HREye devices communicate through \textbf{luceme(s)}, sequences of colored light with semantic meaning.
We next present a sixteen-symbol luceme language with content based on the gestures commonly used for diver-to-diver communication.
Finally, to evaluate the effectiveness of the HREye devices and the lucemes we have defined for them, we present a human study with fourteen participants evaluating the ability of trained participants to recognize active lucemes and the perceived gaze direction of ocular lucemes.
This study demonstrates high levels of accuracy in participant recognition of active lucemes ($83\%$, $92\%$ with high confidence) and reasonable success (21\degree{} avg. error) in the communication of gaze direction using ocular lucemes. 

In this paper, we make the following contributions:
\begin{itemize}
    \item Hardware and software design for HREyes, a new device for light-based AUV-to-human communication.
    \item A sixteen symbol language for the HREyes.
    \item The first method for communicating AUV gaze direction to interactants.
    \item One of the largest in-water human studies of A2H communication, demonstrating the effectiveness of our light-based communication.
        
\end{itemize}
\section{Background: Light-Based Communication, Robot Gaze, and Underwater HRI}
\label{sec:background}

\subsection{Emitted Light For HRI}
\label{sec:background:EL}
Power and status indicator lights have been integrated into the design of robots since the creation of the first robots. 
Modern humanoids such as Pepper~\cite{pandey_mass_2018} and Nao~\cite{han_investigating_2012} often use colored lights around their cameras/eyes to indicate the robot's state, be it power, error, or emotion.
However, the use of light for HRI extends past simple indicator lights.
Light was used as a mediating signal to improve speech interactions by Funakoshi~\etal{}~\cite{funakoshi_smoothing_2008} and used to express emotion by Kim~\etal{}~\cite{kim_determining_2008}.
More recently, Szafir~\etal{}~\cite{szafir_communicating_2015} used light to communicate directionality for aerial robots, while Baraka~\etal{}~\cite{baraka_enhancing_2016, baraka_mobile_2018} developed a variety of ways to use light to communicate state and movement direction for mobile service robots.
Song~\etal{}~\cite{song_bioluminescence_2018, song_effect_2018} applied similar approaches with designs inspired by bioluminescence.

\subsection{Gaze Cues}
\label{sec:background:gaze}
Gaze is one of the most important implicit vectors of communication between humans~\cite{butterworth_what_1991,zeifman_sweet_1996, beier_infants_2012}, used intuitively by people of all ages across all cultures. 
This highlights the importance of gaze for human-robot interaction, particularly in coordinating collaborative work. 
The use of gaze as a vector of implicit communication in human-robot interaction has been well studied~\cite{mavridis_review_2015} for use in establishing shared attention~\cite{scassellati_imitation_1999, scassellati_mechanisms_1996}, managing handovers~\cite{admoni_deliberate_2014, moon_meet_2014}, coordinating task roles~\cite{hoffman_collaboration_2004, breazeal_effects_2005, mutlu_footing_2009}, creating persuasive robots~\cite{chidambaram_designing_2012}, and smoothing social interaction~\cite{breazeal_social_2004}. 
Some robot designers have even explored the development of camera-eye hybrids~\cite{onuki_design_2013, schulz_see_2019} to enable gaze interaction jointly with perception.
Despite the depth of this field, we are unaware of any attempts to provide gaze cues from an AUV, making the ability of the HREye to communicate gaze direction a first for underwater HRI.

\subsection{Underwater HRI and AUV Light Communication}
\label{sec:background:uhri}
In the past two decades, the majority of underwater human-robot interaction (UHRI) publications have focused on enabling human-to-AUV communication with fiducial markers~\cite{dudek_visual_2007}, hand gestures~\cite{xu_natural_2008,islam_dynamic_2018,chiarella_novel_2018}, and remote control~\cite{verzijlenberg_swimming_2010}.
The few works that have explored the inverse question of AUV-to-human communication have primarily focused on the use of digital displays~\cite{dudek_visual_2007, ukai_swimoid_2013}, remote control devices~\cite{verzijlenberg_swimming_2010}, and low-complexity light systems~\cite{verzijlenberg_swimming_2010, demarco_underwater_2014}.
Our previous work in UHRI introduced motion-based communication for AUVs~\cite{fulton_rcvm-icra_2019}, as well as a variety of light-based communication methods, which we briefly survey.

\subsubsection*{Single-Light Signals}
\label{sec:urhi:single}
The first mentions of designing emitted light communication for AUVs use one functional light (comprised of multiple LEDs, but controlled together) as a flashing signal. 
An example is shown in Figure~\ref{fig:led_evolution:single}. 
In Verzijlenberg~\etal{}\cite{verzijlenberg_swimming_2010}, the light is mentioned as an option for confirming that the AUV has received a command, while Demarco~\etal{}\cite{demarco_underwater_2014} design a 4-symbol language similar to Morse code. 
Neither approach was quantitatively evaluated.

\subsubsection*{9-LED Array}
\label{sec:uhri:9LED}
Building on these suggestions, we proposed a communication device based on an array of 9 LEDs (Figure~\ref{fig:led_evolution:nine})~\cite{fulton_rcvm-icra_2019}. 
The goal was to add color and position as aspects of encoding, on top of the flash rates suggested in previous work.
When observed by well-trained participants, the 9-LED system achieved a recognition accuracy of $60\%$ and an average response time of $10$s.

\subsubsection*{3-LED Array}
\label{sec:uhri:3LED}
The idea of our 9-LED system was further refined in a 3-LED version (Figure~\ref{fig:led_evolution:three})~\cite{fulton_rcvm-rss_2022} which uses full RGB diodes, enabling those three lights to produce not only the red, yellow, and green lights of the 9-LED system, but also blues, purples, and many other colors.
With well-trained participants, the 3-LED system achieved an accuracy of $70.4\%$ with an average response time of $28.3$ seconds (inflated by the nature of online administered evaluation).

\section{HREyes: Biomimetic LED Arrays}
\label{sec:system}
Our previous work built upon the suggestions of earlier researchers and demonstrated that light-based communication was possible for AUVs.
However, our previous systems had only achieved $70\%$ accuracy at most, were not intuitive at all, and were difficult to create light displays for, given their limited state space.
To provide collaborative AUVs (co-AUVs) with a highly communicative and more intuitive light-based communication method, we present the \textbf{HREye}, a biomimetic LED device for AUV-to-human communication. 
HREyes are comprised of 40 individually addressable RGB light-emitting diodes arranged in two concentric circles.
Built by daisy-chaining two Adafruit NeoPixel rings, one with 24 LEDs and one with 16 LEDs, an HREye can be used on its own or synchronously with another. 
In our implementation, the HREyes are controlled using the Robot Operating System (ROS) and a microcontroller that drives the NeoPixel rings.
Two HREyes integrated into the LoCO AUV~\cite{edge_design_2020} can be seen in a variety of environments in Figure~\ref{fig:hreye_environments}.

\begingroup
\setlength\tabcolsep{1.5pt}
\renewcommand{\arraystretch}{0}
\begin{figure}[t]
\vspace{2mm}
    \centering
    \adjustbox{max width=0.75\linewidth}{
    \begin{tabular}{cc}
        \raisebox{2\normalbaselineskip}[0pt][0pt]{\rotatebox[origin=l]{90}{\quad\quad\huge\textbf{Lab}}} & \includegraphics[width=\linewidth, trim={0cm 0cm 0cm 1cm}, clip]{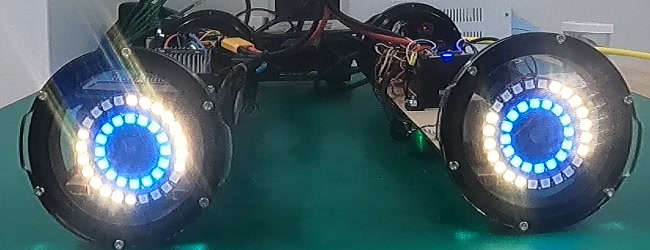}\\
        
        \raisebox{1.7\normalbaselineskip}[0pt][0pt]{\rotatebox[origin=l]{90}{\quad\quad\huge\textbf{Pool}}} & \includegraphics[width=\linewidth, trim={10cm 8cm 19cm 16.5cm}, clip]{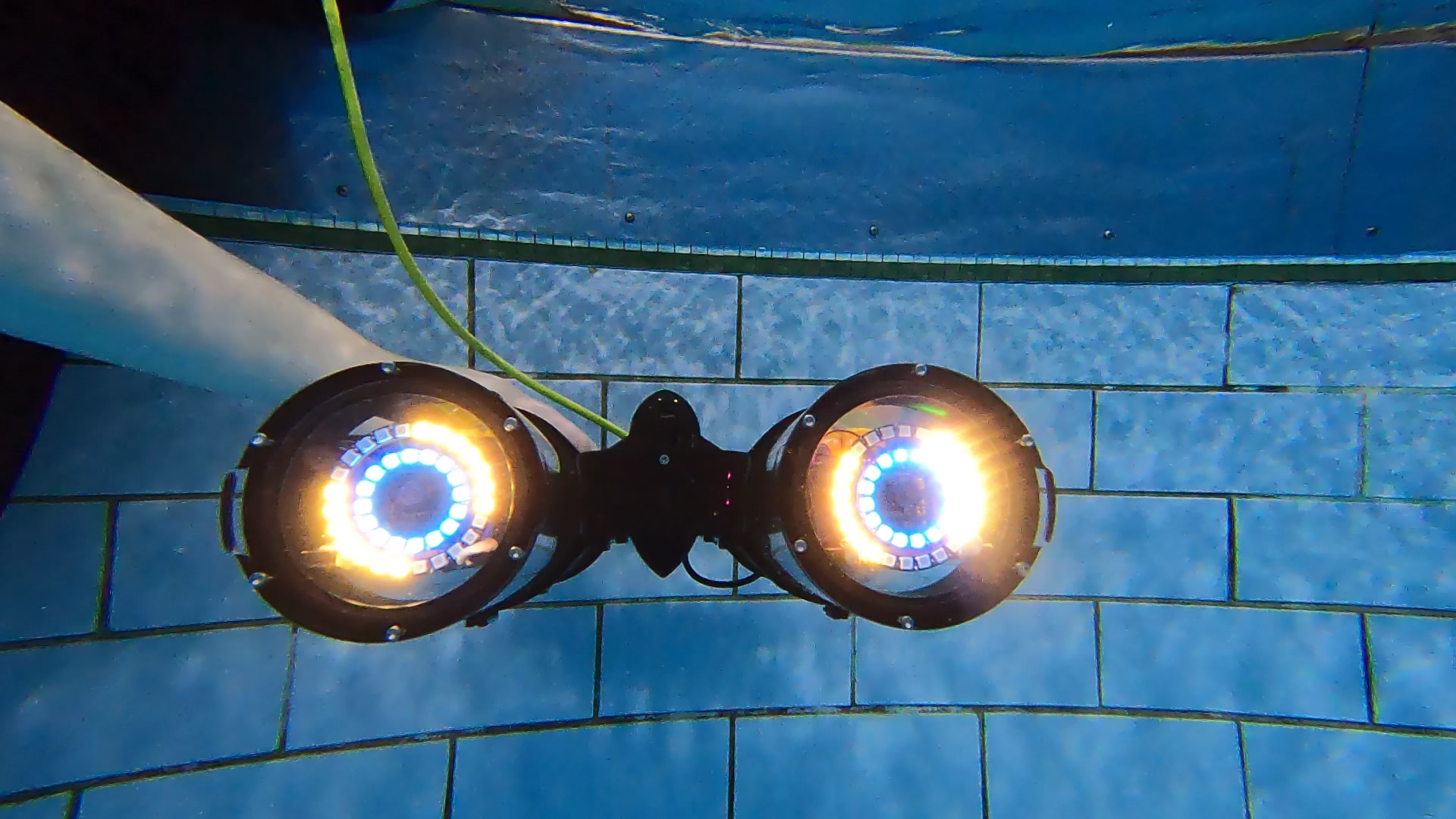}\\
        \raisebox{1.5\normalbaselineskip}[0pt][0pt]{\rotatebox[origin=l]{90}{\quad\quad\huge\textbf{Lake}}} & \includegraphics[width=\linewidth, trim={16cm 11cm 5cm 11cm}, clip]{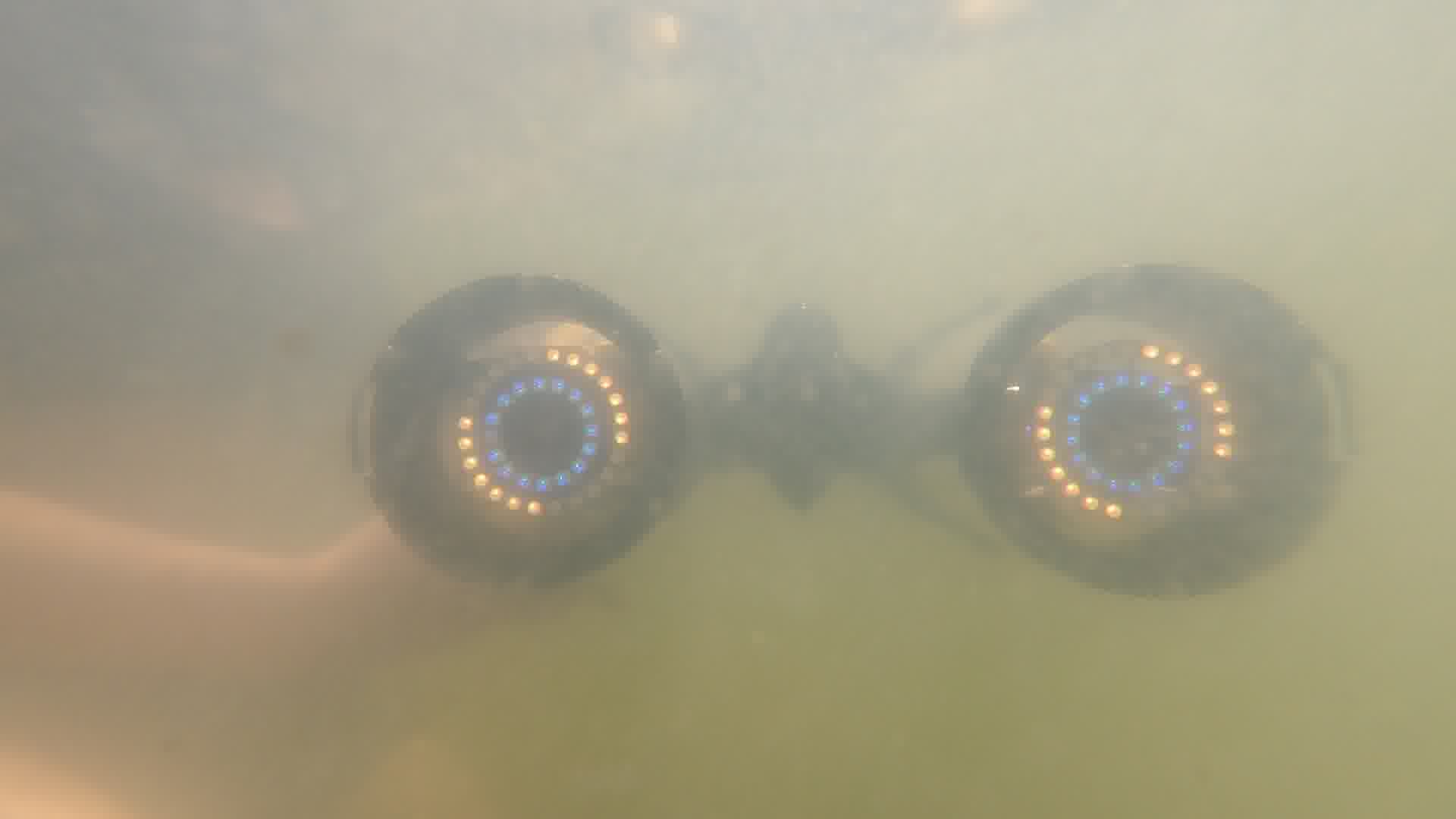}\\
    \end{tabular}}
    \caption{HREyes in LoCO AUV, demonstrating \luceme{FollowYou} in a variety of environments including a laboratory, a pool, and a Minnesota lake.}
    \label{fig:hreye_environments}
    \vspace{-5mm}
\end{figure}
\endgroup

\begingroup
\setlength\tabcolsep{1.5pt}
\renewcommand{\arraystretch}{0}
\begin{figure}[ht]
    \centering
    \adjustbox{max width=\linewidth}{
    \begin{tabular}{|c|c|}
        \hline
         \textbf{Luceme}\textbackslash\textbf{Meaning} & \textbf{Visuals}  \\ \hline
        
         \makecell{\luceme{Affirmative}\\ Yes, Okay. }& 
         \makecell{\centering
         \includegraphics[height=27pt]{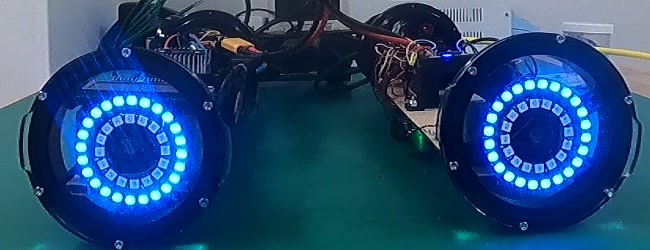}
         \hspace{-2.75mm} 
         \includegraphics[height=27pt]{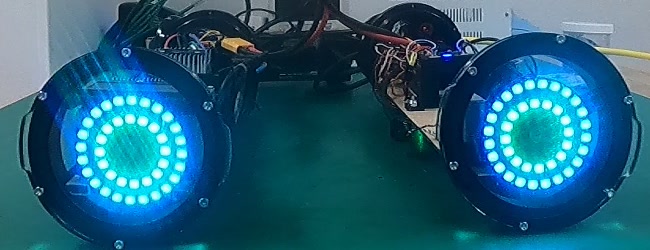} }\\ \hline
         
         \makecell{\luceme{Negative}\\ No. }&
         \makecell{\centering\includegraphics[height=27pt]{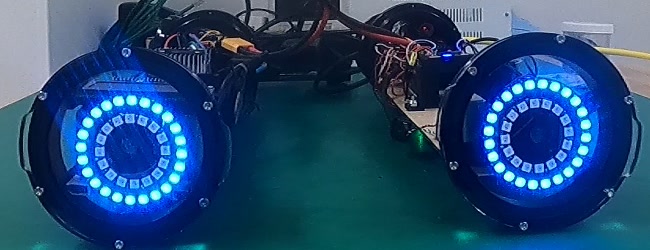} 
         \hspace{-2.75mm} 
         \includegraphics[height=27pt]{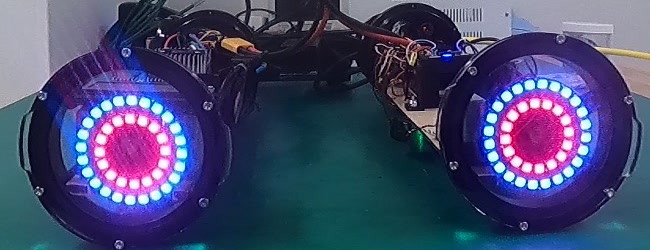}} \\ \hline
         
         \makecell{\luceme{Danger}\\ Danger in the area. }&
         \makecell{\centering\includegraphics[height=27pt]{img/active/lights_off.jpeg}
         \hspace{-2.75mm} 
          \includegraphics[height=27pt]{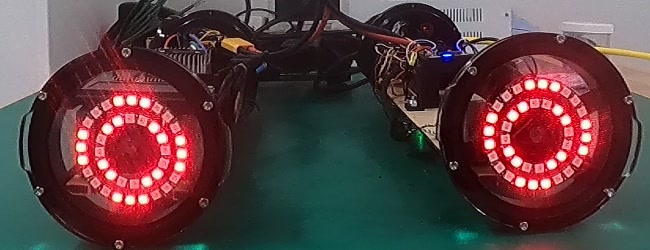}} \\ \hline
         
         \makecell{\luceme{Attention}\\ Pay attention to AUV.} & 
         \makecell{\centering\includegraphics[height=27pt]{img/active/lights_off.jpeg} 
         \hspace{-2.75mm} 
         \includegraphics[height=27pt]{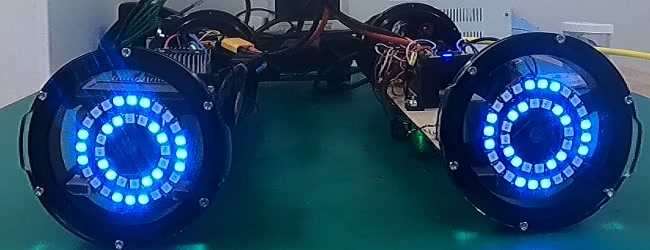} }\\ \hline
         
         \makecell{\luceme{Malfunction}\\ Internal malfunction. }&
         \makecell{\centering\includegraphics[height=27pt]{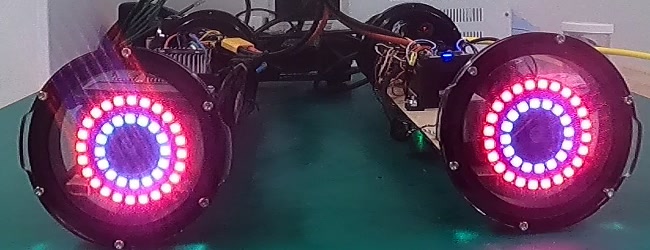} 
         \hspace{-2.75mm} 
         \includegraphics[height=27pt]{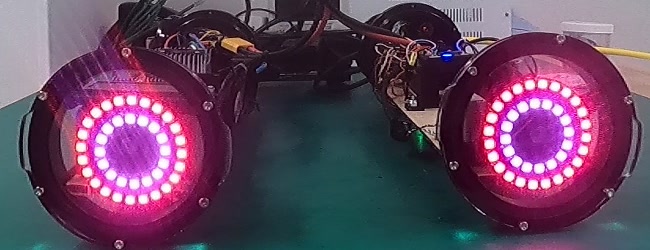}} \\ \hline
         
         \makecell{\luceme{Wait CMD}\\ Waiting for instructions.}&
         \makecell{\centering\includegraphics[height=27pt]{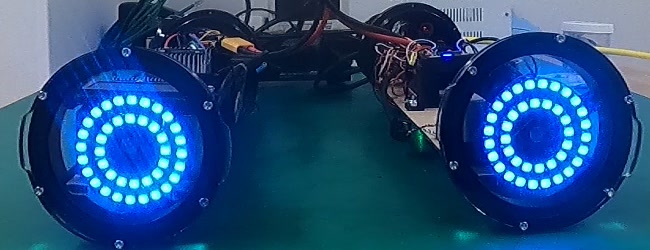} 
         \hspace{-2.75mm} 
         \includegraphics[height=27pt]{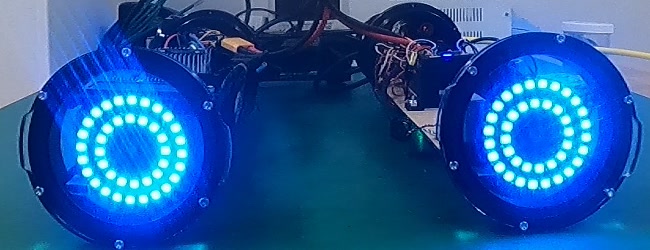}} \\ \hline
         
         \makecell{\luceme{Go Left}\\Go left/AUV going left.}&
         \makecell{\centering\includegraphics[height=27pt]{img/active/lights_off.jpeg} 
         \hspace{-2.75mm} 
         \includegraphics[height=27pt]{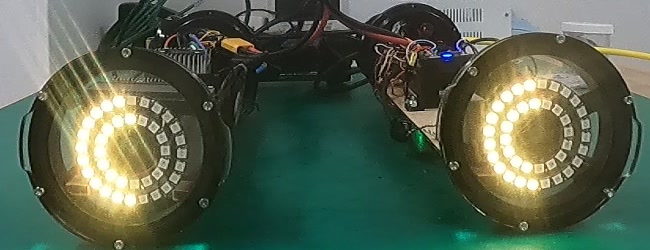}} \\ \hline

         \makecell{\luceme{Go Right}\\Go right/AUV going right.}&
         \makecell{\centering\includegraphics[height=27pt]{img/active/lights_off.jpeg} 
         \hspace{-2.75mm} 
         \includegraphics[height=27pt]{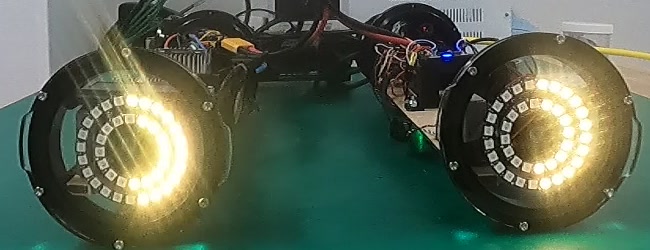}} \\ \hline
         
         \end{tabular}
         \begin{tabular}{|c|c|}
         \hline
         \textbf{Luceme}\textbackslash\textbf{Meaning} & \textbf{Visuals}  \\ \hline
         
         \makecell{\luceme{Go Up}\\Go up/AUV going up.}&
         \makecell{\centering\includegraphics[height=27pt]{img/active/lights_off.jpeg} 
         \hspace{-2.75mm} 
         \includegraphics[height=27pt]{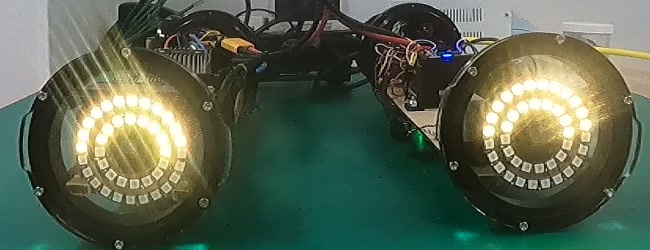}} \\ \hline
         
         \makecell{\luceme{Go Down}\\Go down/AUV going down.}&
         \makecell{\centering\includegraphics[height=27pt]{img/active/lights_off.jpeg} 
         \hspace{-2.75mm} 
         \includegraphics[height=27pt]{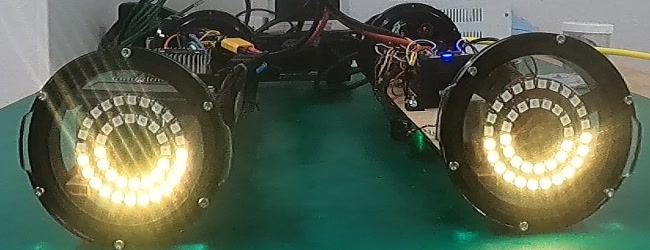}} \\ \hline
         
         \makecell{\luceme{Which Way}\\ Asking for directions.}&
         \makecell{\centering\includegraphics[height=27pt]{img/active/left.jpeg} 
         \hspace{-2.75mm} 
         \includegraphics[height=27pt]{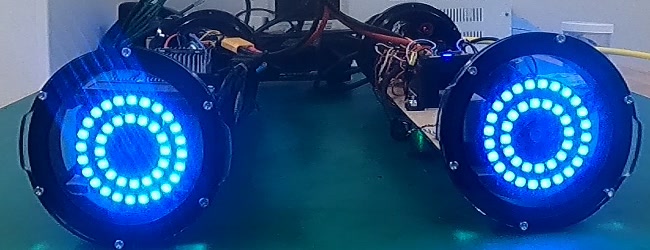}}
         \\ \hline
         
         \makecell{\luceme{Stay}\\ Stay where you are.} & 
         \makecell{\centering\includegraphics[height=27pt]{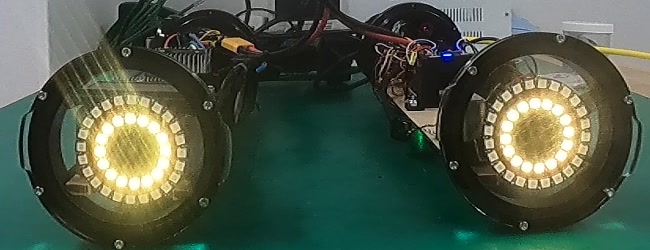}
         \hspace{-2.75mm} 
          \includegraphics[height=27pt]{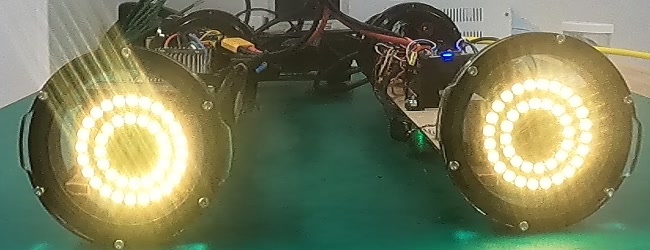}}\\ \hline
         
         \makecell{\luceme{Come Here}\\ Come to the AUV.}&
         \makecell{\centering\includegraphics[height=27pt]{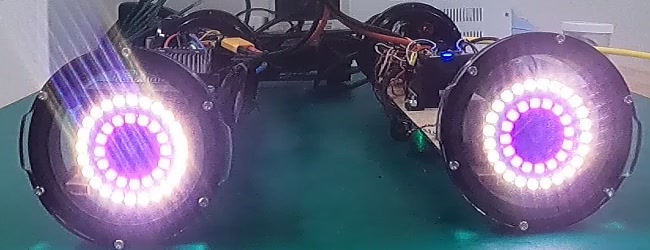} 
         \hspace{-2.75mm} 
         \includegraphics[height=27pt]{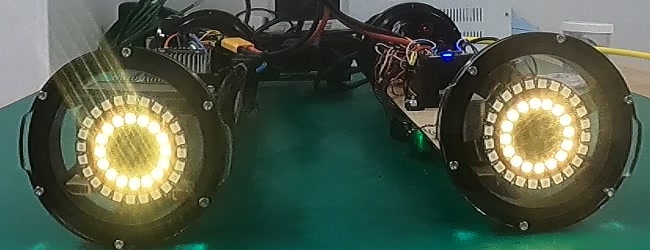}}\\ \hline
         
         \makecell{\luceme{Follow Me}\\Diver can follow AUV.} & 
         \makecell{\centering\includegraphics[height=27pt]{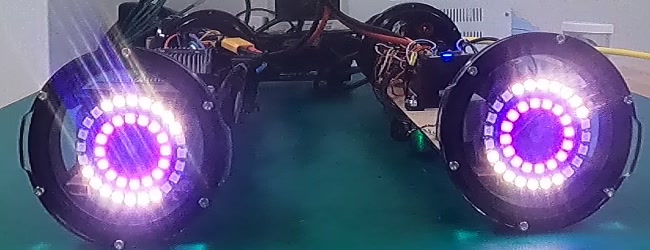} 
         \hspace{-2.75mm} 
         \includegraphics[height=27pt]{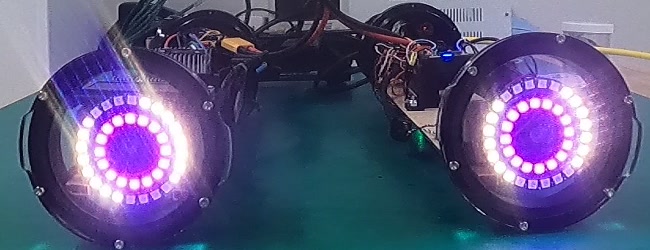}} \\ \hline
         
         \makecell{\luceme{Follow You}\\ AUV will follow diver.}&
         \makecell{\centering\includegraphics[height=27pt]{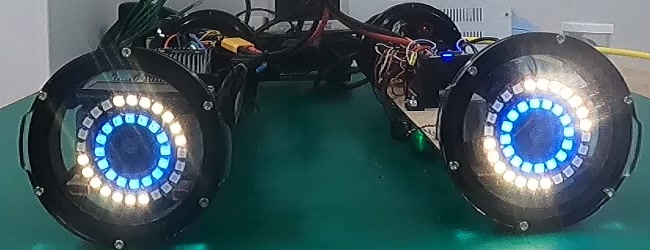} 
         \hspace{-2.75mm} 
         \includegraphics[height=27pt]{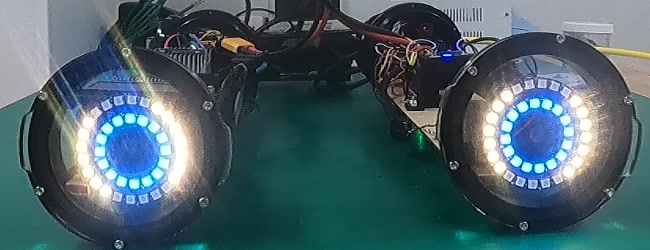}} \\ \hline
         
         \makecell{\luceme{Battery Level}\\ Battery level is...} &
         \makecell{\centering\includegraphics[height=27pt]{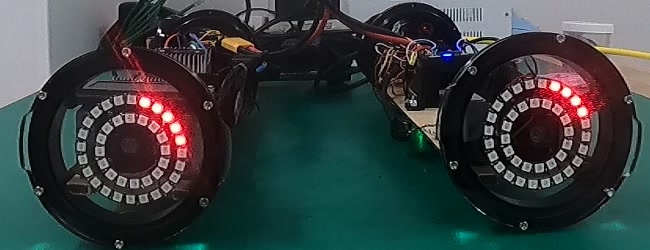} 
         \hspace{-2.75mm} 
         \includegraphics[height=27pt]{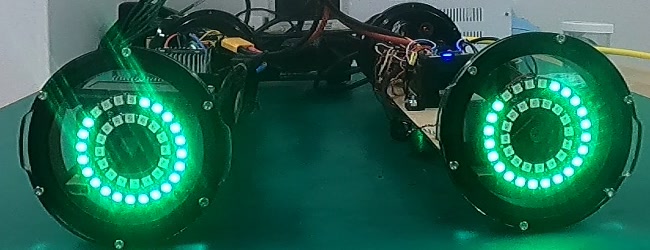}} \\ \hline
    \end{tabular}
    }
    \caption{Selected active lucemes, demonstrated in a laboratory.}
    \label{fig:active_luceme_table}
    \vspace{-5mm}
\end{figure}
\endgroup

\subsection{Design Inspiration}
\label{sec:system:hreye:design}
When designing the HREye, the goal was to create a light-based communication device capable of displaying complex light displays, producing an amount of light useful for illuminating the AUV's environment and indicating direction in an intuitive manner. 
To enable the goal of complex displays, we chose to increase the number of LEDs from previous systems and arrange them in a structure which enables the creation of complex shapes. 
Next, to illuminate the scene for an AUV's cameras, the lights were placed near the AUV's camera. 
Many participants in previous studies have noted that the cameras of AUVs such as LoCO and Aqua seem to be the \enquote{eyes} of the AUV.
This gave rise to the idea for HREyes: arrange a large number of LEDs around the AUV's camera and mimic the appearance of the human eye. 
This helps to fulfill the final design goal by allowing the HREye to mimic human gaze cues. 


\subsection{Control Modalities}
An HREye's state can be represented by a set of 40 tuples with 4 values (red, green, blue, alpha) across time. 
Each tuple maps to a specific LED on the two rings of the HREye. 
Each HREye is controlled by a ROS node (the \textit{hreye\_driver}) running on a microcontroller that receives messages containing an array of 40 tuples, which represent the instantaneous state of HREye.
The microcontroller processes these  messages and applies the appropriate states to the LEDs.
A secondary ROS node (the \textit{hreye\_controller}) is responsible for creating these state vectors. 
The \textit{hreye\_controller} has three modes: active, ocular, and functional.
The lucemes which comprise active and ocular modes will be discussed in the following section.
In functional mode, the HREye device simply acts as a functional light, providing artificial lighting for the AUV at a variety of intensities and color grades.  
This mode is not discussed further, as its uses are purely functional.

\section{Designing Light With Meaning}
\label{sec:lucemes}

\subsection{Definition: Luceme}
In either active or ocular mode, the HREye device is designed to produce animated sequences of colored light called \textbf{lucemes}.
These sequences include the color, order, duration, and in some cases the intensity of illumination. 
The word luceme comes from the Latin \textit{luc}, meaning \enquote{light} combined with the Greek \textit{eme}, a unit of linguistic structure.
A luceme is designated by the symbol \luceme{Meaning}, where \enquote{Meaning} is a simple description of the meaning. 
Active and ocular lucmemes are displayed in Figures~\ref{fig:active_luceme_table} and~\ref{fig:ocular_luceme_table} respectively, but can viewed better in the companion video.

\subsection{Active Lucemes}
\label{sec:lucemes:active}
In active mode, the HREye device produces lucemes with specific semantic meaning. 
The current luceme versions are shown in Figure~\ref{fig:active_luceme_table}. 
The symbols of this language were selected by extracting a set of common communication phrases from diver gestural languages through a process of coding and clustering.
A luceme was designed for each of the concepts extracted from these languages.
These lucemes utilize a color and structure mapping protocol intended to improve recognition accuracy. 
First, for color: yellow relates to directional commands or information, red is connected to problems or danger, blue is mapped to information, and purple is used to reference the AUV itself. 
In terms of structure, there is a shared symbol used for \luceme{Danger} and \luceme{Attention} to indicate time-crucial information, a pulsing illumination used for \luceme{WaitCMD} and \luceme{Malfunction} to indicate a state which requires diver intervention, and a circular animation with one segment of light following another to connect the \luceme{FollowMe} and \luceme{FollowYou} lucemes. 
Together, these color, shape, and motion mappings result in lucemes with related meanings having similar appearances, which allows interactants to more easily memorize them.

\subsection{Ocular Lucemes}
\label{sec:lucemes:ocular}
Ocular lucemes are much more uniform than their active counterparts. 
All ocular lucemes follow the same structure: the inner ring is illuminated in pink (recalling the iris), with the outer ring illuminated in white (recalling the sclera). 
The color pink was selected as it had not been used in many lucemes, was similar to the color used to reference the robot, and would be quite noticeable in the water. 
The ocular lucemes created (depicted in Figure~\ref{fig:ocular_luceme_table}) include a steady state luceme, lucemes for blinking, squinting, and widening eyes, and a set of gaze cues with 30\degree{} granularity.
Only the gaze direction ocular lucemes have been evaluated in human studies.

\section{Study: Human Understanding of Lucemes}
\label{sec:study}
In the following sections, we describe a study that tests the recognition accuracy of active and ocular lucemes by evaluating the recognition of human interactants observing the lucemes in a pool.
This study was approved as human research by the University of Minnesota's Institutional Review Board (reference number: 00015327).

\begingroup
\setlength\tabcolsep{1.5pt}
\renewcommand{\arraystretch}{0}
\begin{figure}[!t]
    \vspace{2mm}
    \centering
    \begin{tabular}{|c|c|}
        \hline
        \textbf{Luceme}\textbackslash\textbf{Meaning} & \textbf{Visuals}  \\ \hline
         
         \makecell{\luceme{Ocular-Blink} \\ Mimicking a blink.} &
         \makecell{\centering
         \includegraphics[height=27.5pt]{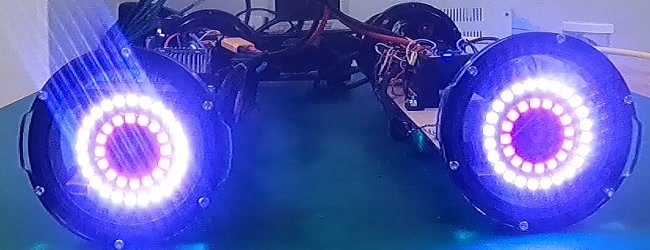} 
         \hspace{-2.75mm} \includegraphics[height=27.5pt]{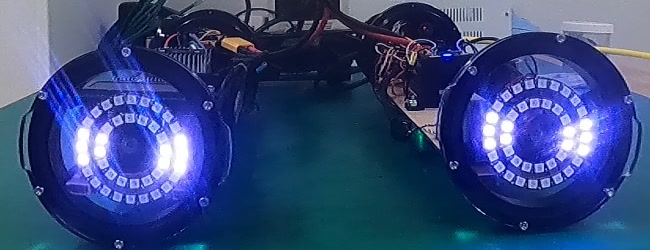} }\\ \hline
         
         \makecell{\luceme{Ocular-Squint} \\ Squinting or focusing.} &
         \makecell{\centering
         \includegraphics[height=27.5pt]{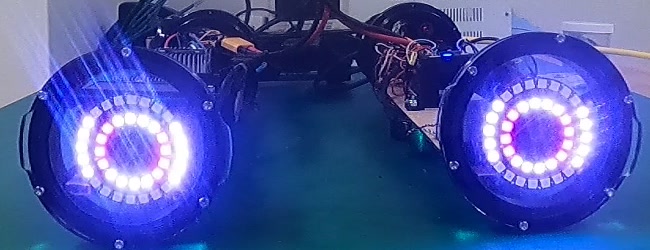} 
         \hspace{-2.75mm} 
         \includegraphics[height=27.5pt]{img/ocular/blink2.jpeg} }\\ \hline
         
         \makecell{\luceme{Ocular-EyesWide} \\ Eyes widen. }& 
         \makecell{\centering
         \includegraphics[height=27.5pt]{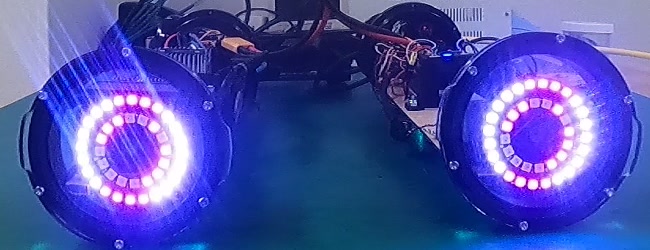} 
         \hspace{-2.75mm} 
         \includegraphics[height=27.5pt]{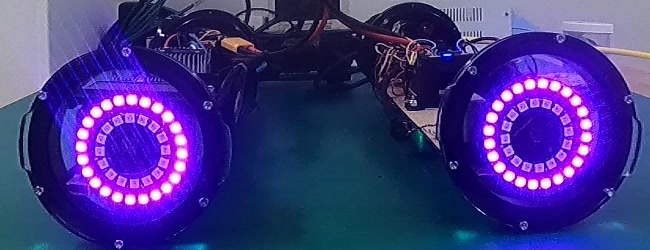} }\\ \hline
         
         \makecell{\luceme{Ocular-90} \\ Gaze cues at 90\degree}. &
         \makecell{\centering
         \includegraphics[height=27.5pt]{img/ocular/ocular_base.jpeg} 
         \hspace{-2.75mm} 
         \includegraphics[height=27.5pt]{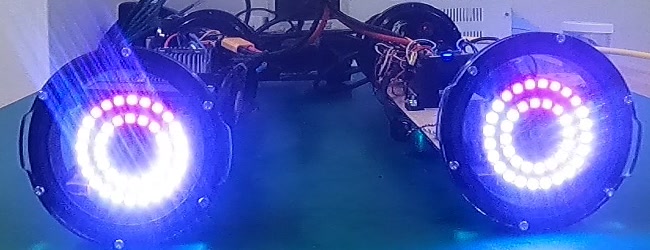} }\\ \hline
         
         \makecell{\luceme{Ocular-120} \\ Gaze cues at 120\degree.} &
         \makecell{\centering
         \includegraphics[height=27.5pt]{img/ocular/ocular_base.jpeg}  
         \hspace{-2.75mm} 
         \includegraphics[height=27.5pt]{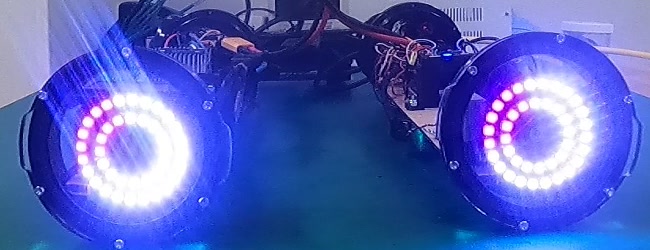} }\\ \hline
    \end{tabular}
    \caption{Selected ocular lucemes, with two examples of directional gaze (angles in Cartesian coordinates).}
    \label{fig:ocular_luceme_table}
    \vspace{-5mm}
\end{figure}
\endgroup

\subsection{Study Design}
\label{sec:study:design}
The study of HREye performance is based on human recognition of HREye lucemes in a pool environment.
To achieve this, we recruited participants, trained them to a point of competency in the use of the HREye system, then asked them to identify active lucemes as produced on the LoCO AUV's HREyes in a pool.
The results from this evaluation serve to demonstrate the performance of the active lucemes of HREyes when being shown to a trained interactant. 
Following this, we also asked each participant to identify the gaze direction of the ocular luceme gaze cues, which they had not been shown previously.
To provide a point of comparison to the HREye-trained participants, we also collected data on human recognition of information passing from the AUV's OLED display.
The participants for this condition were also asked to identify the HREye active and ocular lucemes, providing more data on ocular luceme comprehension along with insights into the ability of untrained participants to recognize active lucemes.

\subsection{Administration Procedures}
\label{sec:study:administration}
\subsubsection{Training}
A total of 14 participants were recruited for the study via student email lists.
After collecting participant consent and administering a short demographic survey, participants were trained to use either the HREye communication device or an OLED display. 
This was done by displaying a video of 4-5 lucemes (or OLED phrases with equivalent meanings) along with their meanings on screen, then testing the participants on the content of the video. 
Once this had been done for all lucemes, a final competency test was administered, asking participants to identify each luceme/OLED phrase. 
Participants who correctly recognized at least 12 out of the 16 phrases shown passed the competency test, and were scheduled for a pool evaluation, which was completed approximately 7 days after training.

\subsubsection{Pool Sessions -- HREye Condition}
In the pool evaluation portion of the survey, participants were given a quick refresher of the list of luceme meanings.
Following this, participants entered the pool approximately $2$ meters from the LoCO AUV and shown the lucemes of the HREye system in a randomized order. 
For each luceme, the participant dove under the water, observed the luceme display, then surfaced and verbally reported the luceme's meaning.
The time from the beginning of the luceme until the beginning of the participant's answer was recorded, along with the participant's confidence in their answer ($0$ - $10$). 
Once all active lucemes had been tested, participants were asked to continue the same process while being shown ocular lucemes for gaze indication (which they had not been shown). 
\subsubsection{Pool Sessions -- OLED Condition}
Participants in the OLED condition followed a similar procedure.
They were first asked to swim the $2$ meter distance to LoCO three times, after which they were asked to observe the OLED displays from LoCO at a closer distance. 
This allows the participant's average swim time to be added to each response, simulating how long it would take the participant to determine the OLED's content if they had to swim to it, as the OLED is unreadable at $2$m.
Most OLED condition participants were then asked to perform the same procedures as the luceme condition participants. 

\begin{figure}
    \vspace{1.5mm}
    \centering
    \includegraphics[width=0.9\linewidth, trim={0.5cm 0cm 1.5cm 0cm}, clip]{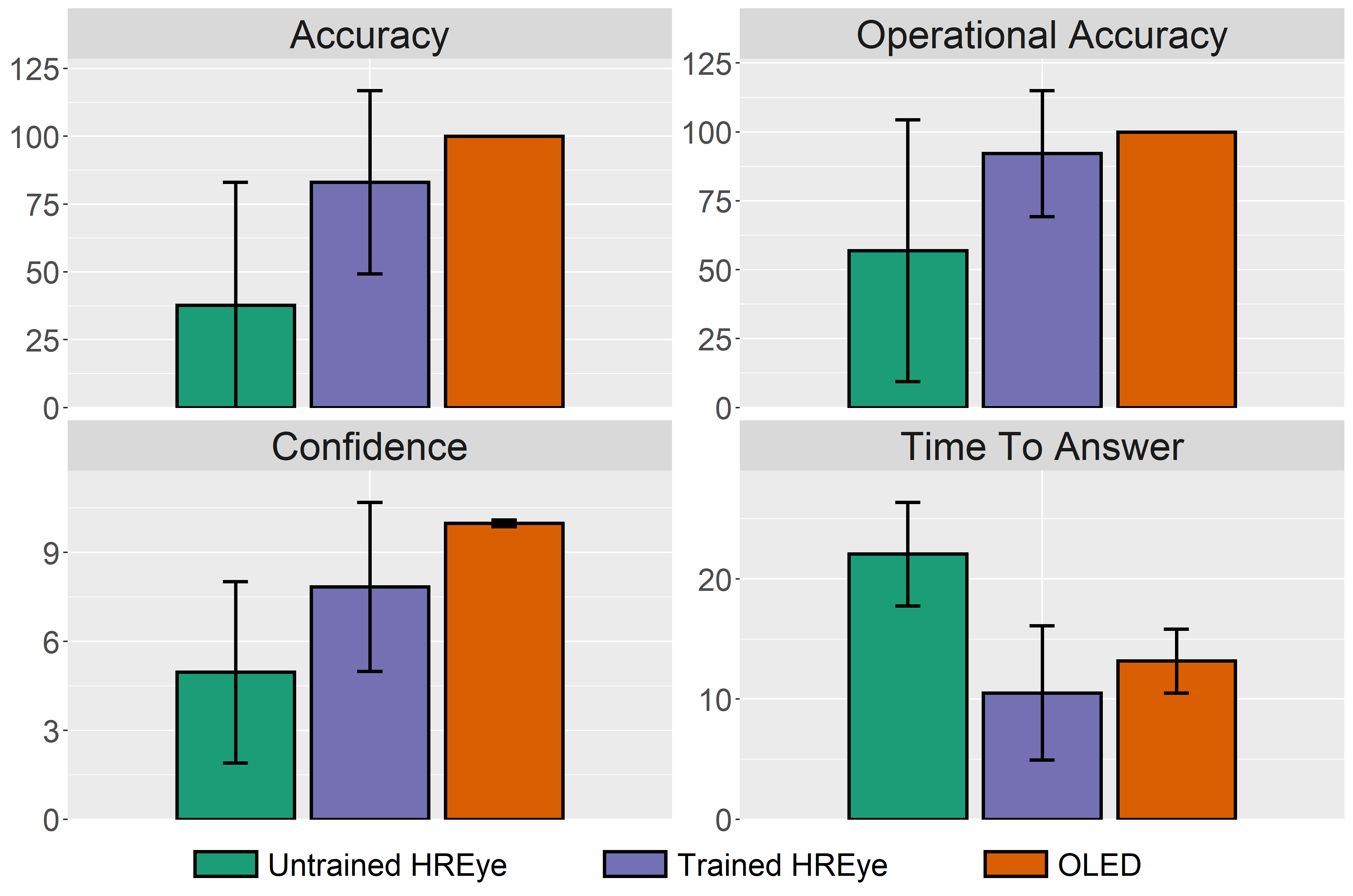}
    \caption{Comparison between the HREye, OLED, and untrained HREye conditions in terms of all metrics.}
    \label{fig:condition_comparison}
    \vspace{-5mm}
\end{figure}

\subsubsection{Debrief Process}
Following their completion of the pool evaluation, participants were asked to complete a quick debrief survey which consisted of a modified Godspeed~\cite{bartneck_measurement_2009, weiss_meta_2015} questionnaire and a set of NASA Task-Load Index (TLX)~\cite{hart_development_1988,noyes_self_2007} questionnaires, one for active luceme/OLED recognition and one for ocular recognition.
The Godspeed questionnaire measures participant opinion on the AUV they were shown, while the NASA TLX questionnaires measure the amount of mental, physical and temporal demand that the luceme recognition placed on participants.
With the debrief survey completed, participants were given a gift card with a value of $10$ USD. 

\subsection{Analysis Procedures}
\label{sec:study:analysis}
Because participants had answered questions about luceme meanings freely, their responses had to be transformed into a quantifiable format for analysis.
For active lucemes, three raters were given a list of active lucemes shown to participants, matched with the meanings participants reported. 
The three raters then scored each pair with a correctness score between $0$ and $100$.
An inter-rater reliability analysis using Fleiss' Kappa~\cite{fleiss_measuring_1971} was performed to evaluate consistency between raters. 
The inter-rater reliability was found to be $\kappa = 0.74$, which is generally taken to mean that there is substantial agreement between raters~\cite{landis_measurement_1977}. 
Since rater consistency was high, the rater's scores were averaged to the final recognition accuracy score.
The raters also transformed ocular luceme responses from various types of input (\eg \enquote{To the upper left}) into an angle of reported gaze ($\kappa = 0.87$).  These translations were also averaged to create a final reported angle.
To complete the analysis, we utilize common metrics in AUV-to-human communication research~\cite{fulton_rcvm-thri_2022}: accuracy (0-100 recognition rate), operational accuracy (0-100, recognition rate of answers with a confidence $\geq 6$), confidence (0-10, participant reported), and time to answer (time between luceme beginning and participant answer).
In the case of OLED participants, the average time it took them to swim to LoCO was added to the time to answer.

\section{Results}
\label{sec:results}
\begin{figure}
    \vspace{2mm}
    \centering
    \includegraphics[width=0.9\linewidth]{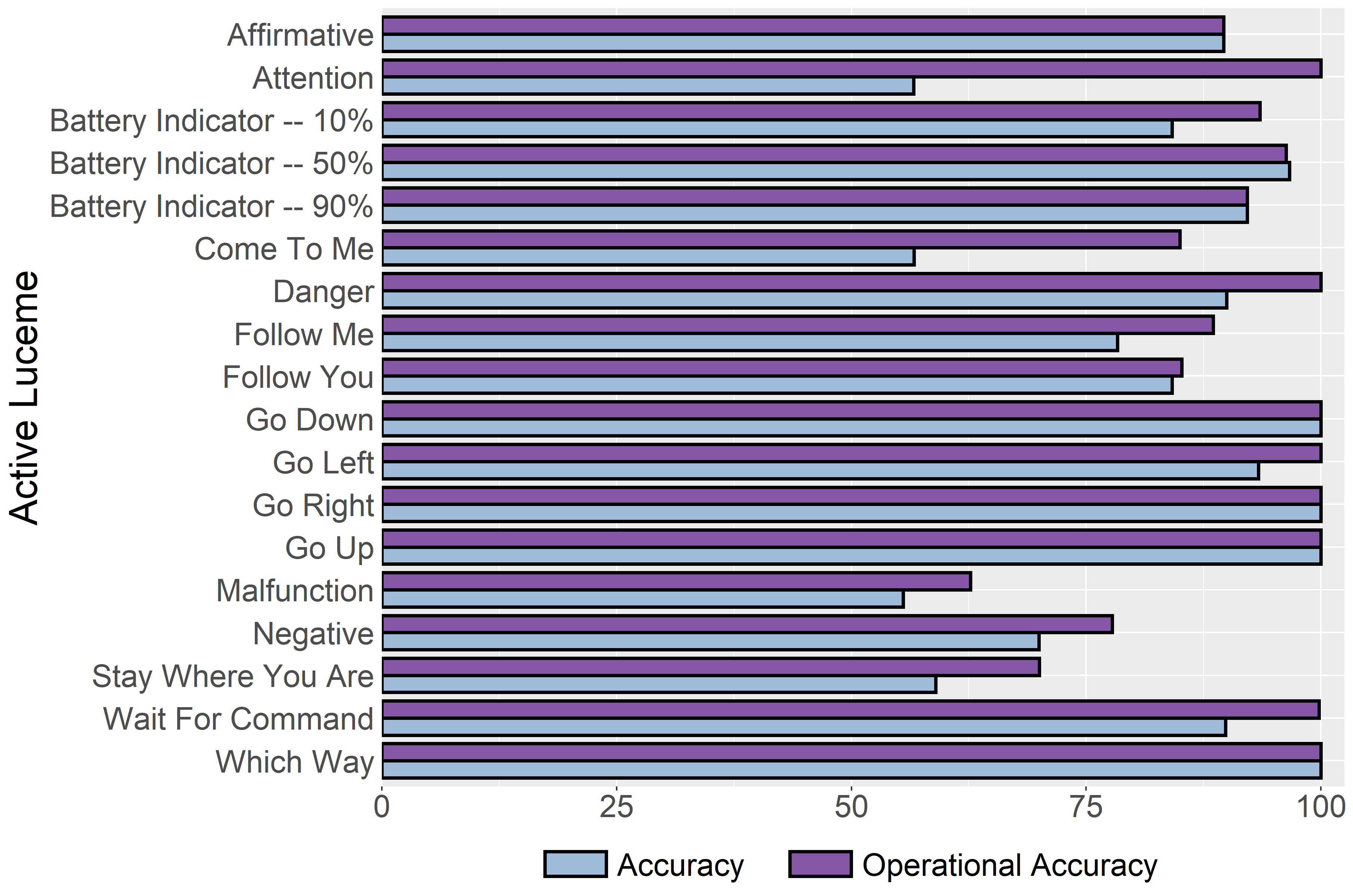}
    \caption{Recognition accuracy and operational accuracy for participants trained to recognize active lucemes.}
    \label{fig:active_paired_accuracies}
    \vspace{-5mm}
\end{figure}

\begin{figure}[ht]
\vspace{2mm}
    \centering
    \includegraphics[width=0.8\linewidth,trim={6cm 0.8cm 6cm 0.9cm},clip]{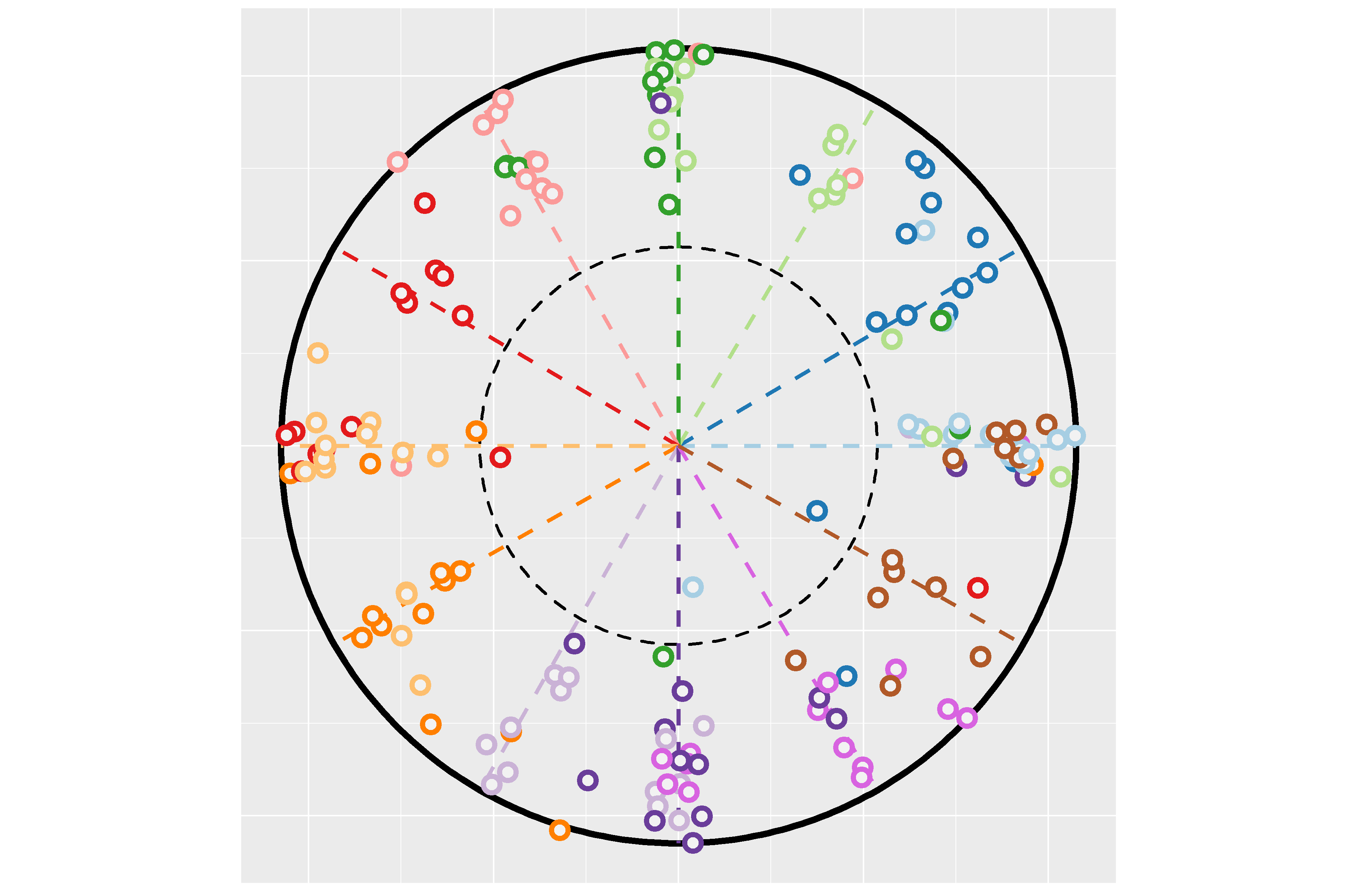}
    \caption{Participant interpretation of gaze lucemes. Points are positioned radially by participant answer and colored by true gaze position. Distance from the center represents participant confidence, with a dashed circle at 5/10 confidence.}
    \label{fig:ocular_accuracy}
    \vspace{-5mm}
\end{figure}

\subsection{Demographics}
\label{sec:results:demo}
Our study population was comprised of 14 people.
Ten participants were trained in the HREye condition, the remaining four were trained on the OLED device, with three of those four also evaluating the HREye system's active lucemes without training.
Ten of these were students, the rest were working or looking for work.
When asked to self-identify their gender, nine identified as male, three as female, and one identified as non-binary.
The majority of participants were between the ages of 18 and 24 and had lived in the US for at least 3 years.
Eight participants indicated that they had experience with robots, four indicated some level of scuba diving experience, and five indicated a familiarity with some kind of sign language.
Nine participants self-reported nearsightedness and none reported colorblindness, which was confirmed to be accurate by a self-administered Ishihara test~\cite{birch_efficiency_1997}.

\subsection{Comparing Across Conditions}
As shown in Figure~\ref{fig:condition_comparison}, the trained HREye participants identified lucemes with a reasonable $83\%$ accuracy. 
This is lower but comparable to the accuracy of OLED trained participants ($100\%$).
The lower accuracy is to be expected as the use of the OLED device requires no learning or memorization, simply the ability to read. 
However, the recognition accuracy of active lucemes, higher than any previous light-based communication system and comparable to that OLED display phrases, supports the idea that the HREye devices could be effectively used in actual deployments.
When considering the accuracy of untrained HREye participants, we see higher accuracy than might perhaps be expected, especially considering previous results on the untrained use of light-based communication systems in ~\cite{fulton_rcvm-icra_2019}.
The accuracy ($37\%$) and operational accuracy ($57\%$) of untrained recognition indicate that HREye lucemes are intuitive to understand.
Further analysis of the performance of specific lucemes indicates that the lucemes most correctly identified by untrained participants were directional lucemes (\luceme{GoLeft}, etc), \luceme{BatteryLevel}, and \luceme{Danger}.

\subsection{Recognition of Active Lucemes}
\label{sec:results:active}
Trained participants identified active lucemes with an overall accuracy of $83\%$.
As shown in Figure~\ref{fig:active_paired_accuracies}, the directional lucemes all achieve high accuracy.
In terms of operational accuracy, also shown in Figure~\ref{fig:active_paired_accuracies}, participants achieve an even higher rate of $92\%$, with a number of lucemes even hitting $100\%$ operational accuracy.
Average participant time-to-answer is $10.5$ seconds, with the \luceme{Attention} and \luceme{Malfunction} being two of the longer response times. 
The level of accuracy (and particularly operational accuracy) demonstrated in this study is sufficient to expect that similar lucemes could be used effectively in the field.

\subsection{Ocular Lucemes}
\label{sec:results:ocular}
Participants correctly identified gaze cues with high accuracy, despite being untrained in this aspect of the system. 
Overall, participants interpreted ocular lucemes with an average error of 21\degree, as visualized in Figure~\ref{fig:ocular_accuracy}.
The actual error is likely lower.
This is due to the fact that in one pool session, one of the HREyes came loose and rotated by almost 45\degree{} before the issue was detected and fixed.
Nonetheless, even with this confounding issue (which affected 3 participants), interactants seem to be able to intuitively understand the gaze direction of an AUV using the HREye ocular lucemes. 
Post-study interviews with some participants suggested that participants did not feel the robot was actually looking in the directions it was indicating, however.
Therefore, it remains to be seen how well ocular lucemes achieve grounding for tasks in underwater environments.

\subsection{Participant Opinion}
In terms of results of the Godspeed and NASA TLX surveys, which ten participants completed, the LoCO AUV was rated as neither particularly anthropomorphic ($1.9$-$2.7$ on average sub-scores out of 5) or animated ($1.8$-$3.0$), but it was considered likable ($3.6$-$4.0$) and intelligent ($3.6$-$4.0$).
The active lucemes were not considered demanding in terms of physical effort ($12.0$ out of 100) or time ($12.9$), but somewhat mentally taxing ($35.1$).
The ocular lucemes were rated at similar levels of physical effort ($9.2$) and time requirements ($14.3$), but lower in terms of mental demand ($19.3$).
\section{Conclusion}
In this paper, we have presented the HREye, a novel, biomimetic, light-based AUV communication device. 
We designed and created a sixteen symbol active luceme language based on diver gestural languages as well as a set of ocular lucemes for communicating gaze direction, and evaluated both types of lucemes in a pool study with fourteen participants.
This study demonstrated high recognition accuracy for active lucemes and intuitive understanding of directional lucemes as well as \luceme{Danger} and \luceme{BatteryLevel}.
Additionally, the ocular lucemes were demonstrated to communicate gaze direction with an average error of 21\degree, a reasonable level of accuracy for the first attempt at gaze indication for AUVs.
These results demonstrate that the HREye device and its active and ocular lucemes are suitable for use in human-robot collaboration, greatly expanding the variety of AUV-to-human communication methods with a novel, robust, and intuitive form of communication.

\bibliographystyle{ieeetr}
\bibliography{refs.bib}
\end{document}